\documentclass{article}

\usepackage[preprint]{neurips_2023}

\usepackage[utf8]{inputenc} % allow utf-8 input
\usepackage[T1]{fontenc}    % use 8-bit T1 fonts
\usepackage{hyperref}       % hyperlinks
\usepackage{url}            % simple URL typesetting
\usepackage{booktabs}       % professional-quality tables
\usepackage{amsfonts}       % blackboard math symbols
\usepackage{nicefrac}       % compact symbols for 1/2, etc.
\usepackage{microtype}      % microtypography
\usepackage{xcolor}         % colors
\usepackage{amsmath}
\usepackage{wrapfig}
\usepackage{graphicx}
\usepackage{multirow}
\usepackage{adjustbox}
\usepackage{multirow}
\usepackage[inline]{enumitem}
\usepackage{colortbl}

\newcommand{\elec}{\cellcolor[HTML]{B9C7FF}}
\newcommand{\stab}{\cellcolor[HTML]{F3DEC6}}
\newcommand{\thermal}{\cellcolor[HTML]{97D58E}}
\newcommand{\optical}{\cellcolor[HTML]{CD888A}}
\newcommand{\mech}{\cellcolor[HTML]{8FD9D4}}

\let\origaddcontentsline\addcontentsline
\def\addcontentsline#1#2#3{}

\title{M$^2$Hub: Unlocking the Potential of Machine Learning \\ for Materials Discovery}

\author{
\normalsize
  Yuanqi Du\textsuperscript{1,*}\quad
  Yingheng Wang\textsuperscript{1,*}\quad
  Yining Huang\textsuperscript{2}\quad
  Jianan Canal Li\textsuperscript{3}\quad
  Yanqiao Zhu\textsuperscript{4}\quad\\
  \normalsize
  \textbf{Tian Xie\textsuperscript{5}\quad
  Chenru Duan\textsuperscript{6}\quad
  John M. Gregoire\textsuperscript{7}\quad
  Carla P. Gomes\textsuperscript{1}}\quad\\
  \normalsize
  \textsuperscript{1} Cornell\quad
  \textsuperscript{2} Northwestern\quad
  \textsuperscript{3} UCB\quad
  \textsuperscript{4} UCLA \quad \\
  \textsuperscript{5} MSR AI4Science \quad
  \textsuperscript{6} Microsoft Quantum\quad
  \textsuperscript{7} Caltech\quad
  \textsuperscript{*} Equal Contribution
}

\begin{document}

\maketitle

\begin{abstract}
  We introduce M$^2$Hub, a toolkit for advancing machine learning in materials discovery. Machine learning has achieved remarkable progress in modeling molecular structures, especially biomolecules for drug discovery. However, the development of machine learning approaches for modeling materials structures lag behind, which is partly due to the lack of an integrated platform that enables access to diverse tasks for materials discovery. To bridge this gap, M$^2$Hub will enable easy access to materials discovery tasks, datasets, machine learning methods, evaluations, and benchmark results that cover the entire workflow. Specifically, the first release of M$^2$Hub focuses on three key stages in materials discovery: virtual screening, inverse design, and molecular simulation, including 9 datasets that covers 6 types of materials with 56 tasks across 8 types of material properties. We further provide 2 synthetic datasets for the purpose of generative tasks on materials. In addition to random data splits, we also provide 3 additional data partitions to reflect the real-world materials discovery scenarios. State-of-the-art machine learning methods (including those are suitable for materials structures but never compared in the literature) are benchmarked on representative tasks. Our codes and library are publicly available at \url{https://github.com/yuanqidu/M2Hub}.
\end{abstract}

\section{Introduction}
\label{sec:intro}

With the methodological advancements in machine learning, an increasing number of machine learning models have been developed and applied to solve scientific problems, from simulating molecular systems with millions of particles to predicting accurate protein structures~\cite{zhang2018deep,jumper2021highly}. 
The primary focus of machine learning in the chemical sciences has remained in the domain of  molecular structures, (bio)molecules including small molecules, proteins, RNAs, etc.~\cite{atz2021geometric,rives2021biological,townshend2021geometric}. However, materials constitute a large portion of the chemical space which have been significantly less studied, especially in the machine learning community.
Among scientific problems, materials discovery plays a vital role in driving innovations and progress across various fields spanning energy, electronics, healthcare, and sustainability~\cite{sanchez2018inverse,gomes2021computational}. However, the traditional trial-and-error approach to materials discovery is expensive and time-consuming. Over decades, classical machine learning methods have already been widely applied in assisting materials discovery,~\cite{schmidt2019recent} yet the impact of machine learning for solid state materials lags behind its efficacy in other areas of chemical science. 

Witnessing the success of machine learning in solving grand challenges in science~\cite{wang2018deepmd,jumper2021highly}, one of the key ingredients is the infrastructure that supports the machine learning community to build the machine learning workflow: data preparation/processing, model development, performance evaluation, and model improvement based on the evaluation feedback. While effort has been made to make materials datasets available to the machine learning community~\cite{blaiszik2019data,dunn2020benchmarking,clement2020benchmark,qayyum2022survey,durdy2023liverpool}, existing work mainly focus on providing data servers that allow the users to query materials data and predefined benchmark sets. However, to bridge the gap between the molecular and solid state materials, we identify the need for a unified platform to facilitate the development of machine learning approaches for materials discovery purpose, including (1) centralized data sources with diverse materials, property and task types, (2) clear problem formulations, (3) realistic problem settings (e.g. data split), and (4) appropriate benchmarks and transparent comparisons with prior methods.

\begin{figure}
\centering
\includegraphics[width=\linewidth]{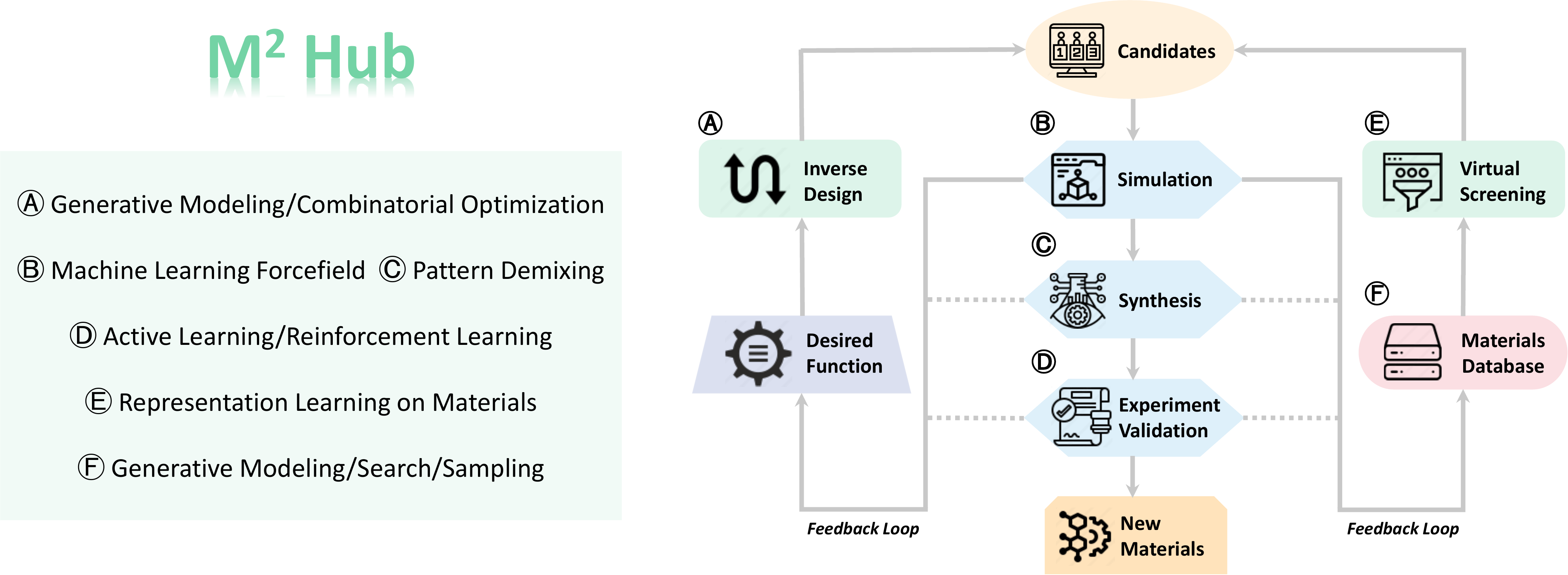}
\caption{M$^2$Hub: \underline{M}aterials discovery meets \underline{M}achine learning. A-F on the left figure demonstrates machine learning approaches used in each stage of the materials discovery pipeline on the right figure (dashed lines denote currently unavailable experiment-related tasks).}
\label{fig:workflow}
\end{figure}

In order to address the aforementioned challenges, we establish M$^2$Hub, which integrates and connects different machine learning building blocks in the materials discovery (Fig.~\ref{fig:workflow}). The cornerstone of M$^2$Hub is a benchmark that incorporates several key aspects: \begin{enumerate*}[label=(\roman*)]
    \item it integrates three key tasks: virtual screening, molecular dynamics simulation, and inverse design, which are translated using machine learning constructs such as materials representation learning, machine learning forcefield, and generative materials design;
    \item it is underpinned by a curated set of 11 datasets that encompass 6 types of materials with 56 tasks across 8 distinct material properties. In addition to the standard random split, we have included 3 realistic (out-of-distribution) data splits to enhance the robustness of model evaluation;
    \item a distinctive feature of our benchmark is the emphasis on the generative design of materials. For this, we provide machine learning formulations, evaluation metrics, and oracle functions to facilitate further research and development in this area;
    \item finally, our benchmarks evaluate not only the commonly used material representation learning methods, but also those designed for non-periodic molecular structures. These methods are applied to 13 representative tasks for material property prediction.
\end{enumerate*}

The flow of this paper is as follows: we introduce the background and related work on developing machine learning methods for materials discovery in Sec.~\ref{sec:background}; we present the overview of M$^2$Hub including problem formulation, dataset curation, data processing, evaluation and oracle function for inverse design in Sec.~\ref{sec:overview}; in Sec.~\ref{sec:benchmark}, we detail the implemented machine learning models, benchmarking results, observations, and insights emerging from the results.

\section{Background}
\label{sec:background}

\subsection{Materials Representation Learning}
Material representation learning refers to representing materials structures in an expressive and machine-readable format for downstream studies, from property prediction to materials generation. Recent advances in graph neural networks bring a wave of representing materials structures as graphs where nodes represent atoms and edges represent bonds or interactions. A line of works has been proposed to adapt this structured inductive bias into deep learning models.CGCNN~\cite{xie2018crystal} introduces a multi-edge graph representation to capture periodicity. MEGNet~\cite{chen2019graph} unifies molecules and crystal structures representations by graph neural networks representing each atom as node and interaction/bond between atoms as edge. More recently, ALIGNN considers both atomistic and line graphs which externally capture angular information~\cite{choudhary2021atomistic}. Equivariant graph neural networks (e.g., E3NN~\cite{geiger2022e3nn}) have also been applied thanks to its roto-translational equivariance property.

\subsection{Machine Learning Forcefields}
Molecular dynamics simulation has become an essential tool to understand the microscopic dynamical behaviors of molecular systems. It is worth noting there is a common trade-off between two popular diagrams, empirical forcefields and \textit{ab initio} molecular dynamics. Empirical forcefields often rely on the hand-crafted parameters which are efficient yet inaccurate, while ab initio molecular dynamics rely on quantum-mechanical calculations which are precise but inefficient.
Inspired by recent advances of deep learning in automated parameters learning and transferability, a large amount of works has been developed to learn machine learning forcefields from quantum-mechanical data to strike a balance between accuracy and efficiency. Specifically, it is expected to be more accurate than empirical forcefields and more efficient than quantum-mechanical calculations. Most representative work include DeepMD~\cite{zhang2018deep}, ANI-1~\cite{smith2017ani}, and NeuqIP~\cite{batzner20223}.

\subsection{Materials Inverse Design}
Designing new materials structures is a long-standing challenge, often known as the inverse design problem, in materials science~\cite{du2022molgensurvey,manica2023accelerating}. Before deep generative models have been applied to this problem, traditional computational methods often leverage quantum mechanical search over the possible stable materials including random search, evolutionary algorithm, element substitutions over known materials~\cite{glass2006uspex,pickard2011ab,hautier2011data}. One line of works leverages a learned force field to minimize the energy of the structure to reach a stable material~\cite{deringer2018data}. Later, deep generative models have been applied to this problem where the models aim to model the distribution of the known crystal structures and learn to sample new structures from it. Early work leverage 3D voxel representation but it is nontrivial to fit atom from the generated voxels~\cite{hoffmann2019data,noh2019inverse}. Later work instead leverage atomic representation directly~\cite{zhao2021high}. G-SchNet~\cite{gebauer2019symmetry} instead proposes an auto-regressive model that generates each atom in a sequential way. Notably, it remains largely unexplored for efficient and controllable crystal structure generation with machine learning methods. A recent work~\cite{xie2022crystal} builds upon the recent success of diffusion models on images and adapts them for crystal structure generation and optimization in an iterative refinement manner instead of one-shot or auto-regressive sequential generation.

\begin{wrapfigure}[16]{r}{0.52\linewidth}
    \vspace{-74 pt}
    \centering
    \includegraphics[width=0.99\linewidth]{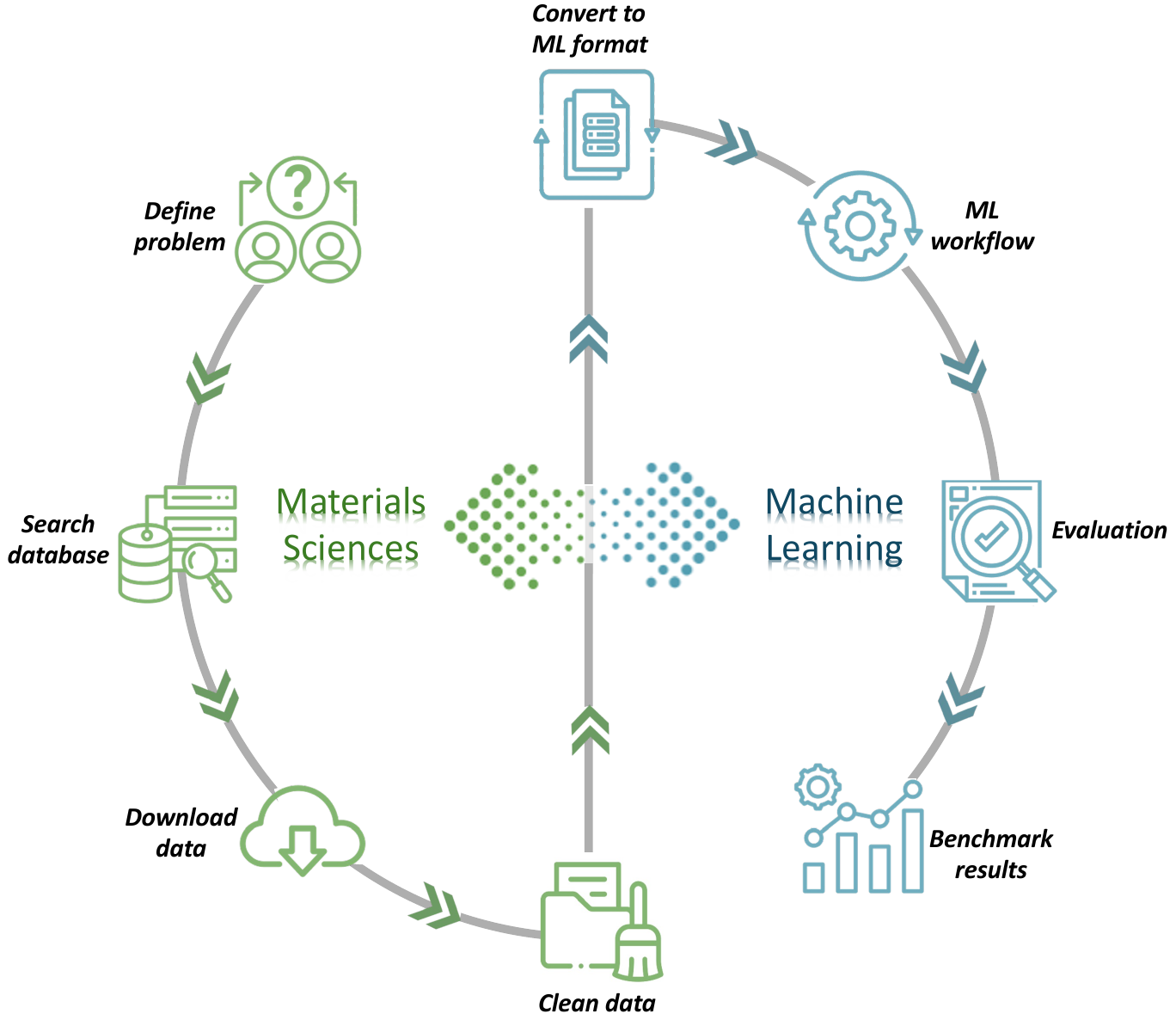}
    \vspace{-3mm}
    \caption{Regular workflow for studying materials with machine learning approaches (green colored steps denote materials science expertise and blue colored steps denote machine learning expertise.}
    \vspace{-4mm}
    \label{fig:pipeline}
\end{wrapfigure}

\section{Overview of M$^2$Hub}
\label{sec:overview}

\subsection{Problem Formulation}
\label{sec:formulation}
\paragraph{Material representation.} Material structures can be represented as a set of atoms $M = (m_0, m_1, \dots, m_N)$ in 3D space with atom types $H = (h_0, h_1, \dots, h_N) \in \mathbb{R}^{N \times K}$ and atomic positions $X = (x_0, x_1, \dots, x_N) \in \mathbb{R}^{N\times 3}$ where $N$ denotes number of atoms and $K$ denotes number of atom types (C, O, Fe, Al, etc.). Most materials are crystal structures which periodically repeat their unit cells in 3D space. In such cases, lattice vectors $L=(l_1, l_2, l_3)\in \mathbb{R}^{3\times 3}$ are utilized to describe the periodicity in 3D space. Note $L$ is not rotation invariant, 6 invariant lattice parameters (lengths of lattice parameters and angles between them) can also be used to describe the lattice $(l_a, l_b, l_c, \alpha, \beta, \gamma)$. Overall, a material is denoted as $M = (H, X, L)$ if it is periodic and otherwise $M = (H, X)$.

\paragraph{Material graph representation.} Regardless of periodicity, materials structures can be naturally represented as graphs $G = (\mathcal{V}, \mathcal{E})$, where $\mathcal{V}$ is a set of vertices and $\mathcal{E}= \{e_{ij}(k_1,k_2,k_3)|i,j\in \{1,2,\dots,N\}, k_1, k_2, k_3 \in \mathbb{Z}\}$ is a set of $D$ edges ($k_1$, $k_2$, $k_3$ denotes the translation of the unit cell using lattice vector $L$, none if not periodic); $H \in \mathbb{R}^{N \times K}$ denotes the node features; $X^{N\times 3}$ denotes the atomic positions; $E \in \mathbb{R}^{D\times F}$ denotes $F$ edge features (such as bonds or distances between each pair of nodes). The graph connections can be determined in multiple ways such as distance threshold, detailed in Sec.~\ref{sec:data_prepare}.

\paragraph{Predictive tasks.} Predictive tasks often have paired input material $M$ and label $Y$, where given an input material $M$, we aim to predict the expected label as $p(Y|M)$. The label $Y$ could be of various format such as binary, scalar, vector and distribution, detailed in Sec.~\ref{sec:eval}. Both material representation learning and machine learning forcefield are considered as predictive tasks.

\paragraph{Generative tasks.} Generative tasks could be divided into two parts: (1) distribution learning and (2) goal-oriented generation. Given a set of $J$ materials $\mathcal{M}=\{M_i\}_{i=1}^{J}$, distribution learning aims to learn the distribution of $p(\mathcal{M})$ and sample new materials $M_{\text{new}}\sim p(\mathcal{M})$. Goal-oriented generation aims to sample molecules fulfilling specific design targets (often defined by an oracle function $f(M)$ such that $M^{\star} = \arg\max_{M \in \mathcal{M}}f(M)$.

\begin{table}
\begin{center}
\caption{Curated materials datasets in M$^2$Hub. Materials type and property type are detailed in Sec.~\ref{sec:mat_type} and Sec.~\ref{sec:prop_type}; dim refers to data dimensionality; PBC refers to perodic boundary condition; method refers to how properties are obtained (sim short for simulation).}
\label{tab:dataset}
\begin{adjustbox}{max width=\textwidth}
\begin{tabular}{ccccccccc}
\toprule
Data name & Materials type & Dim & PBC & Prop. type & Task type & \# tasks & \# data & Method\\
\hline 
MatBench & inorganic bulk & 3D & (T, T, T) & elec./mech./stab./opt/ther. & scalar & 8 & 312--132,752 & Sim. \\
QMOF &  metal-organic framework & 3D & (T, T, T) & elec. & scalar & 1 & >20,000 & Sim. \\
OC20 & bulk-adsorbate interface & 3D & (T, T, F) & energetic & scalar/vector & 3 & 640,081 & Sim.\\
OMDB &  organic crystal & 3D & (T, T, T) & elec. & scalar & 1 & 12,500 & Sim. \\
DFT3D &  inorganic bulk & 3D & (T, T, T) & elec./mech./stab./semi. & scalar & 29 & 55,722 & Sim. \\
DFT2D &  inorganic bulk & 3D & (T, T, F) & stab. & scalar & 1 & 636 & Sim.  \\
EDOS-PDOS &  inorganic bulk & 3D & (T, T, T) & elec./ther. & 1D dist. & 2 & 48,469  & Sim. \\
tmQM & transition metal complex & 3D & (F, F, F) & elec. & scalar & 8 & 86,665 & Sim. \\
QM9 & organic molecules & 3D & (F, F, F) & elec. & scalar & 12 & $\sim$134,000 & Sim. \\
Carbon24 & inorganic bulk & 3D &(T,T,T) & N/A & N/A & 1& 10,153 & N/A\\
Perov5 & inorganic bulk & 3D &(T,T,T) & N/A & N/A & 1& 18,928 & N/A\\
\bottomrule
\end{tabular}
\end{adjustbox}
\end{center}
\end{table}

\subsection{Data Description}

We aim to curate a set of datasets covering the diversity of material, property, and task types and data amounts which will enable a variety of perspectives for machine learning model developments. 

\subsubsection{Material types}
\label{sec:mat_type}

\textbf{Inorganic bulks} refer to solid substances that lacks carbon–hydrogen bonds, that is, substances that are not organic compounds, such as metals, alloys, ceramics, and minerals. They are typically large-scale structures and are used in various applications, ranging from construction materials to electronics, due to their robustness and electrical/thermal conductivity.

\textbf{Organic crystals} are composed of carbon-based molecules arranged in a highly ordered pattern. They exhibit distinct molecular structures and are often used in the field of optoelectronics, such as organic light-emitting diodes (OLEDs), due to their unique optical properties. They can range in size from microscopic to macroscopic.

\textbf{Organic molecules} are individual carbon-based compounds that can be small in size, consisting of a few atoms, or large, such as polymers. They have diverse chemical structures and are widely used in pharmaceuticals, plastics, and organic electronics due to their flexibility in design and functionality.

\textbf{Bulk-adsorbate interfaces} refer to the boundary between a bulk material and an absorbed species, such as gases or liquids, on its surface. They are typically at the nanoscale and play a crucial role in various fields, including catalysis, gas sensing, transport, and energy storage, by influencing the interaction and reactivity of the absorbed species with the material.

\textbf{Transition metal complexes} are coordination compounds consisting of central transition metal atom(s) surrounded by ligands. They exhibit unique electronic and magnetic properties and are commonly used in catalysis, medicine, and material science due to their ability to undergo redox reactions and flexible and tunable coordination environment with organic ligands.

\textbf{Metal-organic frameworks (MOFs)} are crystalline materials composed of metal ions or clusters coordinated with organic linkers. MOFs poss a high surface area and tunable porious structure, making them useful in applications such as gas storage, separations, and catalysis. They can range in size from microscopic crystals to bulk materials.

\subsubsection{Property types}
\label{sec:prop_type}

\textbf{Electrical properties} refer to the characteristics of a material related to its ability to conduct or resist the flow of electric current. These properties include conductivity, resistivity, and dielectric constant, which determine how well a material can conduct or insulate against electrical charges.

\textbf{Mechanical properties} describe how a material behaves under applied forces or loads. These properties include strength, stiffness, ductility, toughness, and elasticity. They determine how the material responds to stress, strain, and deformation, and are essential in understanding its structural integrity and performance.

\textbf{Stability} refers to a material's ability to maintain its properties and resist degradation over time. It encompasses chemical stability (resistance to chemical reactions or corrosion), thermal stability (ability to withstand high temperatures), and mechanical stability (ability to resist physical changes or mechanical stress).

\textbf{Optical properties} pertain to a material's interaction with light. These properties include absorption, reflection, transmission, and emission of light. Optical properties determine a material's color, transparency, opacity, and light-emitting capabilities, and are crucial in fields such as optics, photonics, and display technologies.

\textbf{Thermal properties} describe how a material conducts, stores, and dissipates heat. These properties include thermal conductivity, specific heat capacity, thermal expansion coefficient, and thermal diffusivity. Thermal properties influence a material's ability to transfer heat, its response to temperature changes, and its behavior in thermal management applications.

\textbf{Energetic properties} are computational models that use machine learning algorithms to predict the behavior of materials at the atomic or molecular level. They employ large datasets to learn the relationships between atomic arrangements and energies, enabling the simulation and understanding of complex material systems.

\textbf{Semiconductor properties} refer to the electrical behavior of materials that exhibit an intermediate conductivity between conductors and insulators. These materials can be controlled to selectively allow or impede the flow of electrons, making them ideal for electronic devices. Semiconductor properties are characterized by parameters such as band gap, carrier mobility, and doping concentration, and are crucial in the design and functionality of transistors, diodes, and integrated circuits.

\subsubsection{Datasets}
\label{sec:dataset}

\textbf{Materials Project (MP)}~\cite{jain2013commentary} (license: CC-BY-4.0) is a database that curates inorganic materials with computed properties including but not limited to thermal, electrical, mechanical, etc. 

\textbf{MatBench}~\cite{dunn2020benchmarking} (MIT license) is a benchmark that provides a standardized framework for evaluating and comparing the performance of different machine learning models on various materials science tasks. It curates data from multiple sources with MP as a main source. However, they do not provide machine learning ready data preparation nor implemented machine learning models and workflow.

\textbf{Quantum MOF Database (QMOF)}~\cite{rosen2022high} (license: CC-BY-4.0)  is a comprehensive database that focuses on metal-organic frameworks (MOFs) with quantum-chemical properties. The MOFs are optimized by DFT derived from both experimental and hypothetical MOF databases.

\textbf{Organic Materials Database (OMDB)}~\cite{borysov2017organic} (open access but no license specified) is a repository of organic materials. The properties are calculated using DFT for crystal structures contained in the COD database (in Appendix~\ref{sec:add_data} additional data sources). 

\textbf{Joint Automated Repository for Various Integrated Simulations (JARVIS)}~\cite{choudhary2020joint} (license: GNU v3.0) is a database that integrates materials data from various sources, including quantum mechanical calculations, materials simulations, machine learning predictions and high-throughput databases. Our datasets DFT3D, DFT2D and EDOS-PDOS are all from JARVIS database.

\textbf{Open Catalyst (OC)}~\cite{chanussot2021open} (MIT license) is a database focused on catalytic materials. It includes three tasks: Structure to Energy and Forces (S2EF), Initial Structure to Relaxed Structure (IS2RS) and Relaxed Energy (IS2RE). 

\textbf{Transition Metal Quantum Materials Database (tmQM)}~\cite{balcells2020tmqm} (license: CC BY-NC 4.0) is a comprehensive database focused on transition metal-based materials. It compiles experimentally derived and computationally predicted data on the structure, composition, and electronic properties of transition metal compounds. 

\textbf{Quantum Machines 9} (QM9)~\cite{ramakrishnan2014quantum} (open access but no license specified) comprises small organic molecules up to 9 heavy atoms with 12 quantum chemical properties.

\textbf{Carbon24}~\cite{carbon2020data} (license: CC-BY-4.0) is a synthetic dataset that includes materials made up by carbon atoms but with different structures obtained by \textit{ab initio} random structure searching.

\textbf{Perov5}~\cite{castelli2012new,castelli2012computational} (license: CC-BY-4.0) is a synthetic dataset that includes perovskite materials with the same structure but different compositions.

\subsection{Machine Learning Ready Dataset Preparation}
\label{sec:data_prepare}

\textbf{Raw data format}
The raw data format for both molecules and crystals is 3D structures and atomic types. Other features (such as angular information) can be derived from them. 

\textbf{Machine learning ready data format}
As explained in Sec.~\ref{sec:formulation}, a machine learning ready format for materials includes \textit{atomic types} denote the atomic number of a given atom and are often converted to one-hot embeddings; \textit{atomic coordinates} denote the positions of a given atom and often need to be used careful if equivariance needs to be guaranteed; \textit{edge features} denote information attached to each edge which often include bond types, interatomic distances, etc.

\textbf{Graph construction} Three common graphs are constructed to represent the materials: \textit{multi-graph construction} is a common way to represent materials as graphs which considers edges with repeated atoms (outside the unit cell) as multiple edges with the same atom; \textit{line graph construction} for materials representation is first proposed in~\cite{choudhary2021atomistic} which a bond adjacency graph (i.e. line graph) is constructed to capture the bond and angular information.

\textbf{Data split} The test scenarios are often out-of-distribution of the training set. While previously common use data split is random split, it is crucial to develop data splits that mimic the real scenarios: \textit{composition split} (e.g. AxBy vs AzBy) refers to splitting the dataset with same materials compositions but varying ratios; \textit{system split} (e.g. AB, AC vs ABC) refers to splitting the dataset with unseen materials systems; \textit{time split} refers to splitting the dataset into training, validation and test set by the date when the materials are published. Note that time split is only available for MP dataset now as the publication information for each material structure is provided in MP.

\subsection{Evaluations}
\label{sec:eval}

\subsubsection{Predictive Evaluations}

The evaluation metrics for predictive tasks depend on the type of prediction label: 
\begin{enumerate*}[label=(\roman*)]
    \item \textbf{scalar value prediction}: common evaluation metrics include R$^2$, mean absolute error, and mean squared error;
    \item \textbf{classification}: common evaluation metrics are accuracy and Area under the ROC Curve (AUC-ROC) score;
    \item \textbf{vector/tensor value prediction}: common evaluation metrics are similar to scalar value prediction, including R$^2$, mean squared error, and mean absolute error. However, the distance measurement between two vector or tensor values may need to take into account the symmetry, e.g., rotation invariance, such that the distance between the rotated crystal structure and the original structure is zero;
    \item \textbf{distribution prediction}: common evaluation metrics include cosine similarity, KL divergence, Wasserstein distance, etc.
\end{enumerate*}

\subsubsection{Generative Evaluations}

Evaluating generative tasks has been a notoriously challenging problem in machine learning. The evaluation metrics can generally be divided into three categories: 
\begin{enumerate*}[label=(\roman*)]
    \item \textbf{reconstruction}: This evaluates the performance of the generative methods in reconstructing the exact material in the training set.
    \item \textbf{basic requirement}: This assesses the minimum requirement for the generated materials, such as structure or composition validity.
    \item \textbf{distribution}: This measures whether the generative model is capable of learning the data distribution (in terms of structure, property, etc.) in the training set, and whether it can interpolate or generalize to unseen materials.
\end{enumerate*}

We include all three types of evaluations metrics developed by~\cite{xie2022crystal}: 
\textit{Materials match} is a reconstruction metric to check if the generated material reconstructs structure in the test set. Following~\cite{xie2022crystal}, this is done using \textit{StructureMatcher} from \textit{pymatgen}~\cite{ong2013python} which considers the match of two materials considering invariances. 
\textit{Validity} is a basic metric to check if the generated materials are valid. Following~\cite{court20203}, a material structure is valid if the shortest distance between any pair of atoms is smaller than 0.5\AA. 
\textit{Structure coverage} is a distribution metric to check if the generated material structures cover the training distribution. We follow~\cite{xie2022crystal} to utilize the CrystalNN fingerprint~\cite{zimmermann2020local} and normalized MagPie fingerprint~\cite{ward2016general} to define the structure and composition distance, respectively. 
\textit{Property statistics} is a distribution metric to check if the properties of generated materials are close to those in the training dataset.

\subsection{Oracle Functions for Generative Materials Design}

In our efforts to facilitate the generative design of materials, we have established two categories of oracle functions. Drawing inspiration from oracle functions designed for drug discovery via machine learning~\cite{huang2021therapeutics}, we initially offer a \textbf{fingerprint (FP)-based oracle function}. This function utilizes conventional materials descriptors in tandem with classical machine learning algorithms to predict properties of interest. More specifically, we have pre-trained a random forest model for each material property prediction task across 13 different datasets as proposed by \citet{dunn2020benchmarking}. Consequently, by harnessing the pre-trained models with extracted features based on the Sine Coulomb Matrix and MagPie featurization algorithms, we can predict the properties of an input material. While the predictive accuracy of these classical, FP-based materials descriptors may not rival that of deep learning-based models, we underscore the importance of their inclusion. Their utilization enables generalization where rules apply and mitigates the risk of biasing the optimization process towards deep learning. Our second oracle function is \textbf{structure-based oracle function} which aids in selecting an appropriate substrate for a given material (film). By taking into account their respective structures, we have incorporated an oracle function that matches a film with a list of substrates. Specifically, this method analyzes the compatibility between a thin film and various potential substrates, particularly in terms of crystallographic orientation, matching area, potential strain, and elastic energy. This is achieved by loading the structural information of the film and substrates from respective files, then calculating and grouping matches based on substrate Miller indices. Each match, characterized by a minimum match area, is recorded with relevant details such as the substrate formula, orientations of the film and substrate, and, if available, elastic energy and strain. Then the most suitable matches—those with the smallest matching area—for each substrate orientation are identified. This method ultimately returns a list of all matches, providing a comprehensive overview of how well the film could potentially fit on each substrate. Details can be found in Appendix~\ref{sec:oracle}.

\section{Benchmarking Machine Learning Models}
\label{sec:benchmark}

\subsection{Existing approach}

A burgeoning amount of machine learning models have been developed for learning molecular representations suitable for a variety of downstream tasks, especially machine learning potential and molecular property prediction~\cite{wu2018moleculenet, ramakrishnan2014quantum, chmiela2017machine}. However, most of existing work focus on molecules without periodicity. Around the same time, another branch of work motivated directly by modeling crystal structures have been developed. We implement and benchmark models from both branches to facilitate the development of new methods in realization of both directions. Specifically, we detail them below and summarize them in Table~\ref{tab:method}.

\textbf{Learning on crystal structures.} \textit{CGCNN}~\cite{xie2018crystal} is a E(3) invariant graph neural network that leverages pairwise distances as edge features.
\textit{ALIGNN}~\cite{choudhary2021atomistic} is an E(3) invariant graph neural network that builds an extra line graph to explicitly encode the bond angle information in addition to the original atomistic graph similar to CGCNN.

\textbf{Learning on molecular structures.}
\textit{SchNet}~\cite{schutt2018schnet} is an E(3) invariant graph neural network that leverages pairwise distances with a continuous filter convolution to construct the message. 
\textit{EGNN}~\cite{satorras2021n} is an E(3) equivariant graph neural network that leverages relative positions between each pair of nodes and pairwise distances as the message function to update both invariant and equivariant features.  
\textit{DimeNet++}~\cite{gasteiger2020fast} is an SE(3) invariant graph neural network that introduces bond angles to improve expressiveness. However, it requires triplet of atom representations to model the bond angle.
\textit{GemNet}~\cite{gasteiger2021gemnet} is an SE(3) invariant graph neural network that leverages dihedral angles for better expressiveness. However, it requires learning on quadruplet representations of atoms.
\textit{Equiformer}~\cite{liaoequiformer} is an SE(3)/E(3) equivariant graph transformer network. Equiformer equips previous transformers with equivariant operations such as tensor product to learn equivariant features built from irreducible representations. 
\textit{LEFTNet}~\cite{LEFTNet} is an SE(3)/E(3) equivariant graph neural network based on equivariant local frames. LEFTNet first scalarizes vector and tensor features during message passing and convert them back by tensorizing the scalars through the equivariant frames proposed in ClofNet~\cite{du2022se}. LEFTNet introduces a local structure encoding and frame transition encoding components to further improve the expressiveness.

\begin{table}
\begin{center}
\caption{Representative work in modeling molecular and crystal structures.}
\label{tab:method}
\begin{adjustbox}{max width=\textwidth}
\begin{tabular}{ccccc}
\toprule
Method & Representation &  Symmetry & Graph construction & Angular\\\hline 
CGCNN~\cite{xie2018crystal} & Graph &  Perm. + E(3) Inv. & Multi-graph & None \\
ALIGNN~\cite{choudhary2021atomistic} & Graph &  Perm. + E(3) Inv. & Multi-graph + Line graph & Explicit\\
SchNet~\cite{schutt2018schnet} & Graph & Perm. + E(3) Inv. & Multi-graph & None \\
EGNN~\cite{satorras2021n} & Graph & Perm. + E(3) Equiv. & Multi-graph & Implicit \\\
DimeNet++~\cite{gasteiger2020fast} & Graph & Perm. + SE(3) Inv. & Multi-graph & Explicit\\
GemNet~\cite{gasteiger2021gemnet} & Graph & Perm. + SE(3) Inv. & Multi-graph & Explicit \\
Equiformer~\cite{liaoequiformer} & Graph & Perm. + E(3)/SE(3) Equiv. & Multi-graph & Implicit \\
LEFTNet~\cite{LEFTNet} & Graph & Perm. + E(3)/SE(3) Equiv. & Multi-graph & Implicit \\
\bottomrule
\end{tabular}
\end{adjustbox}
\end{center}
\end{table}

\subsection{Experiment Set-ups}
We build on top of the Open Catalyst Project (OCP) which provides reproducible implementations of commonly used 3D graph neural networks with benchmarks on OC datasets~\cite{chanussot2021open}. We further implement CGCNN, ALIGNN, EGNN, Equiformer and LEFTNet as they are not included in OCP. We test all the methods on a list of 13 representative tasks from our benchmarks with three data splits (random, composition and system). We mostly use the default hyperparameters provided in the open-source code of each method and reported them in Appendix~\ref{sec:exp_detail}. As OC20 and QM9 have been largely adopted in the community, we directly take the results and report in Appendix~\ref{sec:add_results}. Most of our experiments are conducted on single 16GB V100s while some experiments with memory-intensive models on single 80GB A100s.

\subsection{Results and Discussions}
Several observations can be gleaned from our benchmark results as shown in Table~\ref{tab:main_results}: \begin{enumerate*}[label=(\roman*)]
    \item \textbf{performance} \textit{(observation 1)}: despite the competitive performance of advanced equivariant graph neural networks, invariant models such as DimeNet++ and ALIGNN continue to be among the state-of-the-art methods;
    \item \textbf{efficiency} \textit{(observation 2)}: there is a significant variation in efficiency across the benchmarked models. ALIGNN, DimeNet++, GemNet, and Equiformer, as illustrated in Table~\ref{tab:eff}, have particularly slow runtimes. LEFTNet presents a desirable balance of accuracy and efficiency;
    \item \textbf{data split} \textit{(observation 3)}: more realistic data splits indeed increase the challenge of the task, particularly the system split. However, this trend does not hold for all properties, with dielectric being an exception;
    \item \textbf{material type} \textit{(observation 4)}: the performance trends across various models remain consistent for a given material property. For instance, for the bandgap property, organic crystals (OMDB) demonstrate the smallest values, followed by metal-organic frameworks (QMOF), while inorganic bulk materials (MP) exhibit the largest values.
\end{enumerate*}

\begin{table}
\begin{center}
\caption{Benchmark on materials property prediction tasks (different colors denote distinct property types: purple (electrical), yellow (stability), green (thermal), red (optical), blue (mechanical), - denotes missing results due to extremely small test set after data split). The best numbers are darkly shaded, and the second-best numbers are lightly shaded.}
\vspace{-3mm}
\newcommand{\firstplace}{\cellcolor{gray!35}\boldmath}
\newcommand{\secondplace}{\cellcolor{gray!20}}
\label{tab:main_results}
\begin{adjustbox}{max width=\textwidth}
\begin{tabular}{cccccccccccccccc}
\toprule
\multirow{10}{*}{\rotatebox[origin=c]{90}{Random}}  & Methods    & \elec omdb bandgap & \elec qmof bandgap & \elec mp bandgap & \elec is\_metal & \elec edos   & \elec pdos   & \stab dft2d   & \stab perovskites & \stab e\_form & \thermal phonons  & \optical dielectric & \mech log\_gvrh & \mech log\_kvrh & average ranking\\\hline 
& CGCNN      & 0.3351       & 0.2906       & 0.2759     & 89.16\%   & 0.0108 & \secondplace0.0035 & 56.2634 & 0.0600      & 0.0426  & 62.5355  & 0.3122     & 0.0924    & 0.0731   & 5.1 \\
& ALIGNN     & \firstplace{0.2499}       & \firstplace{0.2285}       & \secondplace{0.2010}     & \secondplace90.26\%   & \firstplace0.0104 & 0.0036 & 56.6110 & \firstplace0.0366      & 0.0246  & \firstplace29.4330  & \secondplace0.2827     & \secondplace0.0734    & 0.0557    & \secondplace 2.5 \\

 & SchNet     & 0.4396       & 0.3746       & 0.3298     & 88.45\%   & 0.0120 & 0.0040 & 53.5080 & 0.0713      & 0.0447  & 97.6870  & 0.3295     & 0.1096    & 0.0775  & 7.3  \\
 & EGNN       & 0.5134       & 0.3964       & 0.3263     & 88.15\%   & 0.0113 & \secondplace0.0035 & 51.1220 & 0.0443      & 0.0627  & 65.3510  & 0.3004     & 0.0968    & 0.0735   & 6.2 \\
 & DimeNet++  & 0.2764       & 0.2379       & 0.2062     & \firstplace90.69\%   & 0.0111 & 0.0036 & \firstplace48.5600 & \secondplace0.0371      & \firstplace0.0219  & 39.3630  & \firstplace0.2658     & \firstplace0.0731    & \firstplace0.0522    & \firstplace 2.4 \\
 & GemNet     & \secondplace{0.2516}       & \secondplace{0.2327}       & 0.2014     & 89.75\%   & \firstplace0.0104 & \firstplace0.0034 & 52.5870 & 0.0411      & \secondplace0.0236  & 48.1810  & 0.2985     & 0.0856    & 0.0555   &2.9 \\
 & Equiformer & 0.3900       & 0.3221       & 0.3050     & 88.23\%   & 0.0123 & 0.0039 & 53.4700 & 0.0653      & 0.0568  & 64.7630  & 0.2942     & 0.0983    & 0.0671  &6.2 \\
 & LEFTNet    & 0.3143       & 0.2328       & \firstplace{0.1839}     & 89.96\%   & 0.0110 & 0.0038 & \secondplace49.3590 & 0.0398      & 0.0256  & \secondplace38.0120  & 0.3030     & 0.0792    & \secondplace0.0529  &3.3    \\
\hline
\multirow{8}{*}{\rotatebox[origin=c]{90}{Composition}} 
& CGCNN      & 0.3516       & 0.2965       & 0.2840     & 88.20\%   & 0.0111 & 0.0031 & -       & -           & 0.0411  & -        & 0.3665     & 0.0946    & 0.0610    & 5 \\
 & ALIGNN     & \firstplace{0.2631}       & \firstplace{0.2266}       & \secondplace0.2139     & \secondplace89.00\%   & \firstplace0.0108 & 0.0031 & -       & -           & 0.0249  & -        & 0.3533     & \firstplace0.0761    & 0.0488   &\firstplace 2.3 \\
 & SchNet     & 0.4624       & 0.3670       & 0.3236     & 87.75\%   & 0.0121 & 0.0034 & -       & -           & 0.0432  & -        & 0.3792     & 0.1029    & 0.0600  &6.9  \\
 & EGNN       & 0.5177       & 0.3831       & 0.3433     & 87.31\%   & 0.0117 & \secondplace0.0030 & -       & -           & 0.0625  & -        & 0.3622     & 0.1070    & 0.0685   &7 \\
 & DimeNet++  & 0.2793       & 0.2339       & 0.2234     & \firstplace89.37\%   & 0.0116 & 0.0032 & -       & -           & \firstplace0.0224  & -        & \secondplace0.3303     & \secondplace0.0762    & \firstplace0.0413    &2.7 \\
 & GemNet     & \secondplace{0.2637}       & 0.2321       & 0.2203     & 88.33\%   & \secondplace0.0110 & \firstplace0.0029 & -       & -           & \secondplace0.0245  & -        & \firstplace0.3279     & 0.0807    & \secondplace0.0417   & \secondplace 2.4 \\
 & Equiformer & 0.4050       & 0.3350       & 0.3014     & 87.66\%   & 0.0128 & 0.0032 & -       & -           & 0.0527  & -        & 0.3495     & 0.0938    & 0.0552 & 5.8   \\
 & LEFTNet    & 0.3822       & \secondplace0.2299       & \firstplace0.2062     & 88.79\%   & 0.0116 & 0.0033 & -       & -           & 0.0276  & -        & 0.3515     & 0.0797    & 0.0423  & 3.6  \\ \hline 
\multirow{8}{*}{\rotatebox[origin=c]{90}{System}} & CGCNN      & -            & 0.4224       & 0.6021     & 78.71\%   & 0.0127 & 0.0040 & 53.0300 & 0.1132      & 0.0557  & 142.5260 & 0.1593     & 0.1124    & 0.0966   &5.3  \\
 & ALIGNN     & -            & 0.3431       & \secondplace0.4997     & \secondplace80.07\%   & \secondplace0.0125 & 0.0041 & 43.7480 & 0.0821      & 0.0331  & \firstplace63.4010  & 0.1636     & \firstplace0.0890    & 0.0766   & 3.3 \\
 & SchNet     & -            & 0.4863       & 0.6562     & 76.59\%   & 0.0131 & 0.0044 & 41.1090 & 0.1405      & 0.0512  & 196.0570 & 0.1571     & 0.1194    & 0.0971    & 6.8 \\
 & EGNN       & -            & 0.4923       & 0.7350     & 75.04\%   & 0.0130 & \firstplace0.0038 & \firstplace30.8940 & 0.0981      & 0.0826  & 151.5430 & 0.1495     & 0.1181    & 0.0974  & 5.8   \\
 & DimeNet++  & -            & \secondplace0.3419       & 0.5086     & \firstplace80.87\%   & 0.0130 & 0.0041 & 35.6240 & \secondplace0.0806      & \secondplace0.0294  & \secondplace87.5960  & \firstplace0.1128     & \secondplace0.0893    & 0.0698  & \firstplace 2.3    \\
 & GemNet     & -            & \firstplace0.3412       & 0.5676     & 78.47\%   & \firstplace0.0122 & \secondplace0.0039 & 34.8430 & 0.0851      & 0.0422  & 113.2400 & 0.1383     & 0.0951    & \secondplace0.0712   & 3.2 \\
 & Equiformer & -            & 0.4272       & 0.6381     & 75.02\%   & 0.0136 & 0.0043 & 36.7680 & 0.0955      & 0.0603  & 168.2930 & 0.1375     & 0.1066    & 0.0841  & 5.9 \\
 & LEFTNet    & -            & 0.3468       & \firstplace0.4550     & 78.01\%   & 0.0134 & 0.0043 & \secondplace34.0180 & \firstplace0.0790      & \firstplace0.0288  & 89.1200  & \secondplace0.1277     & 0.0915    & \firstplace0.0710  & \secondplace 3.1 \\

\bottomrule
\end{tabular}
\end{adjustbox}
\end{center}
\end{table}

\begin{table}
\begin{center}
\caption{Benchmark the efficiency of machine learning models with materials in different sizes (pdos$\sim$10, e\_form$\sim$30, qmof$\sim$100) on a single V100 GPU (each row with same batch size except when exceeding the maximum memory, running time for 10 epochs).}
\vspace{-3mm}
\label{tab:eff}
\begin{adjustbox}{max width=\textwidth}
\begin{tabular}{ccccccccc}
\toprule
& CGCNN & ALIGNN & SchNet & EGNN & DimeNet++ & GemNet & Equiformer & LEFTNet\\
\midrule
pdos & 68s & 623s & 77s & 87s & 158s & 203s & 713s & 117s \\
e\_form & 8572s & 41343s & 10589s & 12591s & 35622s & 40801s & 62344s & 13797s\\
qmof bandgap & 678s & 2277s & 512s & 1336s & 5240s & 4572s & 27884s & 2405s\\
average ranking &  1.33 & 6 & 1.67 & 3 & 5.67 & 6 & 8 & 4.33 \\
\bottomrule
\end{tabular}
\end{adjustbox}
\end{center}
\end{table}
\section{Conclusion, Limitation and Future Outlook}
\label{sec:conclusion}

In this paper, we introduce M$^2$Hub as a comprehensive platform for machine learning development in materials discovery. M$^2$Hub is a toolkit that consists of problem formulation, data downloading, data processing, machine learning methods implementations, machine learning training and evaluation procedures, and benchmark results. We cover not only the commonly considered predictive tasks on materials but also provides tools to enable the study of generative tasks on materials. Specifically, we curate 9 datasets constructed by 6 types of materials with 65 tasks across 8 property types for the predictive task. We further provide 2 synthetic datasets for the purpose of generative tasks on materials. We design 3 extra challenging and realistic data split schemes in addition to previously used random split. We believe M$^2$Hub will serve as an essential role in machine learning for materials discovery with datasets, infrastructures and benchmarks.

Despite we formulate the materials discovery pipeline in the machine learning language supported by datasets, infrastructures and benchmarks, most of the tasks do not involve experiments. However, in reality, experiment is the golden standard to test new materials. It remains a challenge to develop datasets and benchmarks for machine learning models to grow in assisting the experiment phase of materials discovery such as phase demixing and experiment planning (related work is summarized in Appendix~\ref{sec:add_related}). Another future direction to extend M$^2$Hub is to improve the usability for materials science community (similar to previous work~\cite{ward2018matminer,jacobs2020materials}), e.g. collect pre-trained models~\cite{xiasystematic,wang2021molecular,wang2021molcloze}, benchmark machine learning models on specific tasks~\cite{kong2022density,bai2023xtal2dos}, etc. Finally, our current benchmark only considers model architectures. However, the training scheme, objective function and task-specific design can be key to solve specific problems.

\bibliographystyle{unsrtnat}
\bibliography{neurips.bib}

\begin{thebibliography}{74}
\providecommand{\natexlab}[1]{#1}
\providecommand{\url}[1]{\texttt{#1}}
\expandafter\ifx\csname urlstyle\endcsname\relax
  \providecommand{\doi}[1]{doi: #1}\else
  \providecommand{\doi}{doi: \begingroup \urlstyle{rm}\Url}\fi

\bibitem[Zhang et~al.(2018)Zhang, Han, Wang, Car, and Weinan]{zhang2018deep}
Linfeng Zhang, Jiequn Han, Han Wang, Roberto Car, and EJPRL Weinan.
\newblock Deep potential molecular dynamics: a scalable model with the accuracy
  of quantum mechanics.
\newblock \emph{Physical review letters}, 120\penalty0 (14):\penalty0 143001,
  2018.

\bibitem[Jumper et~al.(2021)Jumper, Evans, Pritzel, Green, Figurnov,
  Ronneberger, Tunyasuvunakool, Bates, {\v{Z}}{\'\i}dek, Potapenko,
  et~al.]{jumper2021highly}
John Jumper, Richard Evans, Alexander Pritzel, Tim Green, Michael Figurnov,
  Olaf Ronneberger, Kathryn Tunyasuvunakool, Russ Bates, Augustin
  {\v{Z}}{\'\i}dek, Anna Potapenko, et~al.
\newblock Highly accurate protein structure prediction with alphafold.
\newblock \emph{Nature}, 596\penalty0 (7873):\penalty0 583--589, 2021.

\bibitem[Atz et~al.(2021)Atz, Grisoni, and Schneider]{atz2021geometric}
Kenneth Atz, Francesca Grisoni, and Gisbert Schneider.
\newblock Geometric deep learning on molecular representations.
\newblock \emph{Nature Machine Intelligence}, 3\penalty0 (12):\penalty0
  1023--1032, 2021.

\bibitem[Rives et~al.(2021)Rives, Meier, Sercu, Goyal, Lin, Liu, Guo, Ott,
  Zitnick, Ma, et~al.]{rives2021biological}
Alexander Rives, Joshua Meier, Tom Sercu, Siddharth Goyal, Zeming Lin, Jason
  Liu, Demi Guo, Myle Ott, C~Lawrence Zitnick, Jerry Ma, et~al.
\newblock Biological structure and function emerge from scaling unsupervised
  learning to 250 million protein sequences.
\newblock \emph{Proceedings of the National Academy of Sciences}, 118\penalty0
  (15):\penalty0 e2016239118, 2021.
\newblock \doi{10.1073/pnas.2016239118}.
\newblock URL \url{https://www.pnas.org/doi/full/10.1073/pnas.2016239118}.
\newblock bioRxiv 10.1101/622803.

\bibitem[Townshend et~al.(2021)Townshend, Eismann, Watkins, Rangan, Karelina,
  Das, and Dror]{townshend2021geometric}
Raphael~JL Townshend, Stephan Eismann, Andrew~M Watkins, Ramya Rangan, Maria
  Karelina, Rhiju Das, and Ron~O Dror.
\newblock Geometric deep learning of rna structure.
\newblock \emph{Science}, 373\penalty0 (6558):\penalty0 1047--1051, 2021.

\bibitem[Sanchez-Lengeling and Aspuru-Guzik(2018)]{sanchez2018inverse}
Benjamin Sanchez-Lengeling and Al{\'a}n Aspuru-Guzik.
\newblock Inverse molecular design using machine learning: Generative models
  for matter engineering.
\newblock \emph{Science}, 361\penalty0 (6400):\penalty0 360--365, 2018.

\bibitem[Gomes et~al.(2021)Gomes, Fink, van Dover, and
  Gregoire]{gomes2021computational}
Carla~P Gomes, Daniel Fink, R~Bruce van Dover, and John~M Gregoire.
\newblock Computational sustainability meets materials science.
\newblock \emph{Nature Reviews Materials}, 6\penalty0 (8):\penalty0 645--647,
  2021.

\bibitem[Schmidt et~al.(2019)Schmidt, Marques, Botti, and
  Marques]{schmidt2019recent}
Jonathan Schmidt, M{\'a}rio~RG Marques, Silvana Botti, and Miguel~AL Marques.
\newblock Recent advances and applications of machine learning in solid-state
  materials science.
\newblock \emph{npj Computational Materials}, 5\penalty0 (1):\penalty0 83,
  2019.

\bibitem[Wang et~al.(2018)Wang, Zhang, Han, and Weinan]{wang2018deepmd}
Han Wang, Linfeng Zhang, Jiequn Han, and E~Weinan.
\newblock Deepmd-kit: A deep learning package for many-body potential energy
  representation and molecular dynamics.
\newblock \emph{Computer Physics Communications}, 228:\penalty0 178--184, 2018.

\bibitem[Blaiszik et~al.(2019)Blaiszik, Ward, Schwarting, Gaff, Chard, Pike,
  Chard, and Foster]{blaiszik2019data}
Ben Blaiszik, Logan Ward, Marcus Schwarting, Jonathon Gaff, Ryan Chard, Daniel
  Pike, Kyle Chard, and Ian Foster.
\newblock A data ecosystem to support machine learning in materials science.
\newblock \emph{MRS Communications}, 9\penalty0 (4):\penalty0 1125--1133, 2019.

\bibitem[Dunn et~al.(2020)Dunn, Wang, Ganose, Dopp, and
  Jain]{dunn2020benchmarking}
Alexander Dunn, Qi~Wang, Alex Ganose, Daniel Dopp, and Anubhav Jain.
\newblock Benchmarking materials property prediction methods: the matbench test
  set and automatminer reference algorithm.
\newblock \emph{npj Computational Materials}, 6\penalty0 (1):\penalty0 138,
  2020.

\bibitem[Clement et~al.(2020)Clement, Kauwe, and Sparks]{clement2020benchmark}
Conrad~L Clement, Steven~K Kauwe, and Taylor~D Sparks.
\newblock Benchmark aflow data sets for machine learning.
\newblock \emph{Integrating Materials and Manufacturing Innovation},
  9:\penalty0 153--156, 2020.

\bibitem[Qayyum et~al.(2022)Qayyum, Kim, Bong, Chi, and Choi]{qayyum2022survey}
Faiza Qayyum, Do-Hyeun Kim, Seon-Jong Bong, Su-Young Chi, and Yo-Han Choi.
\newblock A survey of datasets, preprocessing, modeling mechanisms, and
  simulation tools based on ai for material analysis and discovery.
\newblock \emph{Materials}, 15\penalty0 (4):\penalty0 1428, 2022.

\bibitem[Durdy et~al.(2023)Durdy, Hargreaves, Dennison, Wagg, Moran, Newnham,
  Gaultois, Rosseinsky, and Dyer]{durdy2023liverpool}
Samantha Durdy, Cameron~J Hargreaves, Mark Dennison, Benjamin Wagg, Michael
  Moran, Jon~A Newnham, Michael~W Gaultois, Matthew~J Rosseinsky, and Matthew
  Dyer.
\newblock The liverpool materials discovery server: A suite of computational
  tools for the collaborative discovery of materials.
\newblock 2023.

\bibitem[Xie and Grossman(2018)]{xie2018crystal}
Tian Xie and Jeffrey~C Grossman.
\newblock Crystal graph convolutional neural networks for an accurate and
  interpretable prediction of material properties.
\newblock \emph{Physical review letters}, 120\penalty0 (14):\penalty0 145301,
  2018.

\bibitem[Chen et~al.(2019)Chen, Ye, Zuo, Zheng, and Ong]{chen2019graph}
Chi Chen, Weike Ye, Yunxing Zuo, Chen Zheng, and Shyue~Ping Ong.
\newblock Graph networks as a universal machine learning framework for
  molecules and crystals.
\newblock \emph{Chemistry of Materials}, 31\penalty0 (9):\penalty0 3564--3572,
  2019.

\bibitem[Choudhary and DeCost(2021)]{choudhary2021atomistic}
Kamal Choudhary and Brian DeCost.
\newblock Atomistic line graph neural network for improved materials property
  predictions.
\newblock \emph{npj Computational Materials}, 7\penalty0 (1):\penalty0 185,
  2021.

\bibitem[Geiger and Smidt(2022)]{geiger2022e3nn}
Mario Geiger and Tess Smidt.
\newblock e3nn: Euclidean neural networks.
\newblock \emph{arXiv preprint arXiv:2207.09453}, 2022.

\bibitem[Smith et~al.(2017)Smith, Isayev, and Roitberg]{smith2017ani}
Justin~S Smith, Olexandr Isayev, and Adrian~E Roitberg.
\newblock Ani-1: an extensible neural network potential with dft accuracy at
  force field computational cost.
\newblock \emph{Chemical science}, 8\penalty0 (4):\penalty0 3192--3203, 2017.

\bibitem[Batzner et~al.(2022)Batzner, Musaelian, Sun, Geiger, Mailoa,
  Kornbluth, Molinari, Smidt, and Kozinsky]{batzner20223}
Simon Batzner, Albert Musaelian, Lixin Sun, Mario Geiger, Jonathan~P Mailoa,
  Mordechai Kornbluth, Nicola Molinari, Tess~E Smidt, and Boris Kozinsky.
\newblock E (3)-equivariant graph neural networks for data-efficient and
  accurate interatomic potentials.
\newblock \emph{Nature communications}, 13\penalty0 (1):\penalty0 1--11, 2022.

\bibitem[Du et~al.(2022{\natexlab{a}})Du, Fu, Sun, and Liu]{du2022molgensurvey}
Yuanqi Du, Tianfan Fu, Jimeng Sun, and Shengchao Liu.
\newblock Molgensurvey: A systematic survey in machine learning models for
  molecule design.
\newblock \emph{arXiv preprint arXiv:2203.14500}, 2022{\natexlab{a}}.

\bibitem[Manica et~al.(2023)Manica, Born, Cadow, Christofidellis, Dave, Clarke,
  Teukam, Giannone, Hoffman, Buchan, et~al.]{manica2023accelerating}
Matteo Manica, Jannis Born, Joris Cadow, Dimitrios Christofidellis, Ashish
  Dave, Dean Clarke, Yves Gaetan~Nana Teukam, Giorgio Giannone, Samuel~C
  Hoffman, Matthew Buchan, et~al.
\newblock Accelerating material design with the generative toolkit for
  scientific discovery.
\newblock \emph{npj Computational Materials}, 9\penalty0 (1):\penalty0 69,
  2023.

\bibitem[Glass et~al.(2006)Glass, Oganov, and Hansen]{glass2006uspex}
Colin~W Glass, Artem~R Oganov, and Nikolaus Hansen.
\newblock Uspex—evolutionary crystal structure prediction.
\newblock \emph{Computer physics communications}, 175\penalty0
  (11-12):\penalty0 713--720, 2006.

\bibitem[Pickard and Needs(2011)]{pickard2011ab}
Chris~J Pickard and RJ~Needs.
\newblock Ab initio random structure searching.
\newblock \emph{Journal of Physics: Condensed Matter}, 23\penalty0
  (5):\penalty0 053201, 2011.

\bibitem[Hautier et~al.(2011)Hautier, Fischer, Ehrlacher, Jain, and
  Ceder]{hautier2011data}
Geoffroy Hautier, Chris Fischer, Virginie Ehrlacher, Anubhav Jain, and Gerbrand
  Ceder.
\newblock Data mined ionic substitutions for the discovery of new compounds.
\newblock \emph{Inorganic chemistry}, 50\penalty0 (2):\penalty0 656--663, 2011.

\bibitem[Deringer et~al.(2018)Deringer, Pickard, and
  Cs{\'a}nyi]{deringer2018data}
Volker~L Deringer, Chris~J Pickard, and G{\'a}bor Cs{\'a}nyi.
\newblock Data-driven learning of total and local energies in elemental boron.
\newblock \emph{Physical review letters}, 120\penalty0 (15):\penalty0 156001,
  2018.

\bibitem[Hoffmann et~al.(2019)Hoffmann, Maestrati, Sawada, Tang, Sellier, and
  Bengio]{hoffmann2019data}
Jordan Hoffmann, Louis Maestrati, Yoshihide Sawada, Jian Tang, Jean~Michel
  Sellier, and Yoshua Bengio.
\newblock Data-driven approach to encoding and decoding 3-d crystal structures.
\newblock \emph{arXiv preprint arXiv:1909.00949}, 2019.

\bibitem[Noh et~al.(2019)Noh, Kim, Stein, Sanchez-Lengeling, Gregoire,
  Aspuru-Guzik, and Jung]{noh2019inverse}
Juhwan Noh, Jaehoon Kim, Helge~S Stein, Benjamin Sanchez-Lengeling, John~M
  Gregoire, Alan Aspuru-Guzik, and Yousung Jung.
\newblock Inverse design of solid-state materials via a continuous
  representation.
\newblock \emph{Matter}, 1\penalty0 (5):\penalty0 1370--1384, 2019.

\bibitem[Zhao et~al.(2021)Zhao, Al-Fahdi, Hu, Siriwardane, Song, Nasiri, and
  Hu]{zhao2021high}
Yong Zhao, Mohammed Al-Fahdi, Ming Hu, Edirisuriya~MD Siriwardane, Yuqi Song,
  Alireza Nasiri, and Jianjun Hu.
\newblock High-throughput discovery of novel cubic crystal materials using deep
  generative neural networks.
\newblock \emph{Advanced Science}, 8\penalty0 (20):\penalty0 2100566, 2021.

\bibitem[Gebauer et~al.(2019)Gebauer, Gastegger, and
  Sch{\"u}tt]{gebauer2019symmetry}
Niklas Gebauer, Michael Gastegger, and Kristof Sch{\"u}tt.
\newblock Symmetry-adapted generation of 3d point sets for the targeted
  discovery of molecules.
\newblock \emph{Advances in neural information processing systems}, 32, 2019.

\bibitem[Xie et~al.(2022)Xie, Fu, Ganea, Barzilay, and
  Jaakkola]{xie2022crystal}
Tian Xie, Xiang Fu, Octavian-Eugen Ganea, Regina Barzilay, and Tommi~S
  Jaakkola.
\newblock Crystal diffusion variational autoencoder for periodic material
  generation.
\newblock In \emph{International Conference on Learning Representations}, 2022.

\bibitem[Jain et~al.(2013)Jain, Ong, Hautier, Chen, Richards, Dacek, Cholia,
  Gunter, Skinner, Ceder, et~al.]{jain2013commentary}
Anubhav Jain, Shyue~Ping Ong, Geoffroy Hautier, Wei Chen, William~Davidson
  Richards, Stephen Dacek, Shreyas Cholia, Dan Gunter, David Skinner, Gerbrand
  Ceder, et~al.
\newblock Commentary: The materials project: A materials genome approach to
  accelerating materials innovation.
\newblock \emph{APL materials}, 1\penalty0 (1):\penalty0 011002, 2013.

\bibitem[Rosen et~al.(2022)Rosen, Fung, Huck, O’Donnell, Horton, Truhlar,
  Persson, Notestein, and Snurr]{rosen2022high}
Andrew~S Rosen, Victor Fung, Patrick Huck, Cody~T O’Donnell, Matthew~K
  Horton, Donald~G Truhlar, Kristin~A Persson, Justin~M Notestein, and
  Randall~Q Snurr.
\newblock High-throughput predictions of metal--organic framework electronic
  properties: theoretical challenges, graph neural networks, and data
  exploration.
\newblock \emph{npj Computational Materials}, 8\penalty0 (1):\penalty0 112,
  2022.

\bibitem[Borysov et~al.(2017)Borysov, Geilhufe, and
  Balatsky]{borysov2017organic}
Stanislav~S Borysov, R~Matthias Geilhufe, and Alexander~V Balatsky.
\newblock Organic materials database: An open-access online database for data
  mining.
\newblock \emph{PloS one}, 12\penalty0 (2):\penalty0 e0171501, 2017.

\bibitem[Choudhary et~al.(2020)Choudhary, Garrity, Reid, DeCost, Biacchi,
  Hight~Walker, Trautt, Hattrick-Simpers, Kusne, Centrone,
  et~al.]{choudhary2020joint}
Kamal Choudhary, Kevin~F Garrity, Andrew~CE Reid, Brian DeCost, Adam~J Biacchi,
  Angela~R Hight~Walker, Zachary Trautt, Jason Hattrick-Simpers, A~Gilad Kusne,
  Andrea Centrone, et~al.
\newblock The joint automated repository for various integrated simulations
  (jarvis) for data-driven materials design.
\newblock \emph{npj computational materials}, 6\penalty0 (1):\penalty0 173,
  2020.

\bibitem[Chanussot et~al.(2021)Chanussot, Das, Goyal, Lavril, Shuaibi, Riviere,
  Tran, Heras-Domingo, Ho, Hu, et~al.]{chanussot2021open}
Lowik Chanussot, Abhishek Das, Siddharth Goyal, Thibaut Lavril, Muhammed
  Shuaibi, Morgane Riviere, Kevin Tran, Javier Heras-Domingo, Caleb Ho, Weihua
  Hu, et~al.
\newblock Open catalyst 2020 (oc20) dataset and community challenges.
\newblock \emph{Acs Catalysis}, 11\penalty0 (10):\penalty0 6059--6072, 2021.

\bibitem[Balcells and Skjelstad(2020)]{balcells2020tmqm}
David Balcells and Bastian~Bjerkem Skjelstad.
\newblock tmqm dataset—quantum geometries and properties of 86k transition
  metal complexes.
\newblock \emph{Journal of chemical information and modeling}, 60\penalty0
  (12):\penalty0 6135--6146, 2020.

\bibitem[Ramakrishnan et~al.(2014)Ramakrishnan, Dral, Rupp, and
  Von~Lilienfeld]{ramakrishnan2014quantum}
Raghunathan Ramakrishnan, Pavlo~O Dral, Matthias Rupp, and O~Anatole
  Von~Lilienfeld.
\newblock Quantum chemistry structures and properties of 134 kilo molecules.
\newblock \emph{Scientific data}, 1\penalty0 (1):\penalty0 1--7, 2014.

\bibitem[Pickard(2020)]{carbon2020data}
Chris~J. Pickard.
\newblock Airss data for carbon at 10gpa and the c+n+h+o system at 1gpa, 2020.
\newblock URL \url{https://archive.materialscloud.org/record/2020.0026/v1}.

\bibitem[Castelli et~al.(2012{\natexlab{a}})Castelli, Landis, Thygesen, Dahl,
  Chorkendorff, Jaramillo, and Jacobsen]{castelli2012new}
Ivano~E Castelli, David~D Landis, Kristian~S Thygesen, S{\o}ren Dahl,
  Ib~Chorkendorff, Thomas~F Jaramillo, and Karsten~W Jacobsen.
\newblock New cubic perovskites for one-and two-photon water splitting using
  the computational materials repository.
\newblock \emph{Energy \& Environmental Science}, 5\penalty0 (10):\penalty0
  9034--9043, 2012{\natexlab{a}}.

\bibitem[Castelli et~al.(2012{\natexlab{b}})Castelli, Olsen, Datta, Landis,
  Dahl, Thygesen, and Jacobsen]{castelli2012computational}
Ivano~E Castelli, Thomas Olsen, Soumendu Datta, David~D Landis, S{\o}ren Dahl,
  Kristian~S Thygesen, and Karsten~W Jacobsen.
\newblock Computational screening of perovskite metal oxides for optimal solar
  light capture.
\newblock \emph{Energy \& Environmental Science}, 5\penalty0 (2):\penalty0
  5814--5819, 2012{\natexlab{b}}.

\bibitem[Ong et~al.(2013)Ong, Richards, Jain, Hautier, Kocher, Cholia, Gunter,
  Chevrier, Persson, and Ceder]{ong2013python}
Shyue~Ping Ong, William~Davidson Richards, Anubhav Jain, Geoffroy Hautier,
  Michael Kocher, Shreyas Cholia, Dan Gunter, Vincent~L Chevrier, Kristin~A
  Persson, and Gerbrand Ceder.
\newblock Python materials genomics (pymatgen): A robust, open-source python
  library for materials analysis.
\newblock \emph{Computational Materials Science}, 68:\penalty0 314--319, 2013.

\bibitem[Court et~al.(2020)Court, Yildirim, Jain, and Cole]{court20203}
Callum~J Court, Batuhan Yildirim, Apoorv Jain, and Jacqueline~M Cole.
\newblock 3-d inorganic crystal structure generation and property prediction
  via representation learning.
\newblock \emph{Journal of Chemical Information and Modeling}, 60\penalty0
  (10):\penalty0 4518--4535, 2020.

\bibitem[Zimmermann and Jain(2020)]{zimmermann2020local}
Nils~ER Zimmermann and Anubhav Jain.
\newblock Local structure order parameters and site fingerprints for
  quantification of coordination environment and crystal structure similarity.
\newblock \emph{RSC advances}, 10\penalty0 (10):\penalty0 6063--6081, 2020.

\bibitem[Ward et~al.(2016)Ward, Agrawal, Choudhary, and
  Wolverton]{ward2016general}
Logan Ward, Ankit Agrawal, Alok Choudhary, and Christopher Wolverton.
\newblock A general-purpose machine learning framework for predicting
  properties of inorganic materials.
\newblock \emph{npj Computational Materials}, 2\penalty0 (1):\penalty0 1--7,
  2016.

\bibitem[Huang et~al.(2021)Huang, Fu, Gao, Zhao, Roohani, Leskovec, Coley,
  Xiao, Sun, and Zitnik]{huang2021therapeutics}
Kexin Huang, Tianfan Fu, Wenhao Gao, Yue Zhao, Yusuf Roohani, Jure Leskovec,
  Connor Coley, Cao Xiao, Jimeng Sun, and Marinka Zitnik.
\newblock Therapeutics data commons: Machine learning datasets and tasks for
  drug discovery and development.
\newblock \emph{Advances in neural information processing systems}, 2021.

\bibitem[Wu et~al.(2018)Wu, Ramsundar, Feinberg, Gomes, Geniesse, Pappu,
  Leswing, and Pande]{wu2018moleculenet}
Zhenqin Wu, Bharath Ramsundar, Evan~N Feinberg, Joseph Gomes, Caleb Geniesse,
  Aneesh~S Pappu, Karl Leswing, and Vijay Pande.
\newblock Moleculenet: a benchmark for molecular machine learning.
\newblock \emph{Chemical science}, 9\penalty0 (2):\penalty0 513--530, 2018.

\bibitem[Chmiela et~al.(2017)Chmiela, Tkatchenko, Sauceda, Poltavsky,
  Sch{\"u}tt, and M{\"u}ller]{chmiela2017machine}
Stefan Chmiela, Alexandre Tkatchenko, Huziel~E Sauceda, Igor Poltavsky,
  Kristof~T Sch{\"u}tt, and Klaus-Robert M{\"u}ller.
\newblock Machine learning of accurate energy-conserving molecular force
  fields.
\newblock \emph{Science advances}, 3\penalty0 (5):\penalty0 e1603015, 2017.

\bibitem[Sch{\"u}tt et~al.(2018)Sch{\"u}tt, Sauceda, Kindermans, Tkatchenko,
  and M{\"u}ller]{schutt2018schnet}
Kristof~T Sch{\"u}tt, Huziel~E Sauceda, P-J Kindermans, Alexandre Tkatchenko,
  and K-R M{\"u}ller.
\newblock Schnet--a deep learning architecture for molecules and materials.
\newblock \emph{The Journal of Chemical Physics}, 148\penalty0 (24):\penalty0
  241722, 2018.

\bibitem[Satorras et~al.(2021)Satorras, Hoogeboom, and Welling]{satorras2021n}
V{\i}ctor~Garcia Satorras, Emiel Hoogeboom, and Max Welling.
\newblock E (n) equivariant graph neural networks.
\newblock In \emph{International conference on machine learning}, pages
  9323--9332. PMLR, 2021.

\bibitem[Gasteiger et~al.(2020)Gasteiger, Giri, Margraf, and
  G{\"u}nnemann]{gasteiger2020fast}
Johannes Gasteiger, Shankari Giri, Johannes~T Margraf, and Stephan
  G{\"u}nnemann.
\newblock Fast and uncertainty-aware directional message passing for
  non-equilibrium molecules.
\newblock \emph{arXiv preprint arXiv:2011.14115}, 2020.

\bibitem[Gasteiger et~al.(2021)Gasteiger, Becker, and
  G{\"u}nnemann]{gasteiger2021gemnet}
Johannes Gasteiger, Florian Becker, and Stephan G{\"u}nnemann.
\newblock Gemnet: Universal directional graph neural networks for molecules.
\newblock \emph{Advances in Neural Information Processing Systems},
  34:\penalty0 6790--6802, 2021.

\bibitem[Liao and Smidt()]{liaoequiformer}
Yi-Lun Liao and Tess Smidt.
\newblock Equiformer: Equivariant graph attention transformer for 3d atomistic
  graphs.

\bibitem[Du et~al.(2023)Du, Du, Wang, Feng, Wang, Ji, Gomes, and Ma]{LEFTNet}
Weitao Du, Yuanqi Du, Limei Wang, Dieqiao Feng, Guifeng Wang, Shuiwang Ji,
  Carla Gomes, and Zhi-Ming Ma.
\newblock Efficient and expressive equivariant graph neural networks.
\newblock \emph{under review}, 2023.

\bibitem[Du et~al.(2022{\natexlab{b}})Du, Zhang, Du, Meng, Chen, Zheng, Shao,
  and Liu]{du2022se}
Weitao Du, He~Zhang, Yuanqi Du, Qi~Meng, Wei Chen, Nanning Zheng, Bin Shao, and
  Tie-Yan Liu.
\newblock Se (3) equivariant graph neural networks with complete local frames.
\newblock In \emph{International Conference on Machine Learning}, pages
  5583--5608. PMLR, 2022{\natexlab{b}}.

\bibitem[Ward et~al.(2018)Ward, Dunn, Faghaninia, Zimmermann, Bajaj, Wang,
  Montoya, Chen, Bystrom, Dylla, et~al.]{ward2018matminer}
Logan Ward, Alexander Dunn, Alireza Faghaninia, Nils~ER Zimmermann, Saurabh
  Bajaj, Qi~Wang, Joseph Montoya, Jiming Chen, Kyle Bystrom, Maxwell Dylla,
  et~al.
\newblock Matminer: An open source toolkit for materials data mining.
\newblock \emph{Computational Materials Science}, 152:\penalty0 60--69, 2018.

\bibitem[Jacobs et~al.(2020)Jacobs, Mayeshiba, Afflerbach, Miles, Williams,
  Turner, Finkel, and Morgan]{jacobs2020materials}
Ryan Jacobs, Tam Mayeshiba, Ben Afflerbach, Luke Miles, Max Williams, Matthew
  Turner, Raphael Finkel, and Dane Morgan.
\newblock The materials simulation toolkit for machine learning (mast-ml): An
  automated open source toolkit to accelerate data-driven materials research.
\newblock \emph{Computational Materials Science}, 176:\penalty0 109544, 2020.

\bibitem[Xia et~al.()Xia, Zhu, Du, and Li]{xiasystematic}
Jun Xia, Yanqiao Zhu, Yuanqi Du, and Stan~Z Li.
\newblock A systematic survey of chemical pre-trained models.

\bibitem[Wang et~al.(2021{\natexlab{a}})Wang, Min, Shao, and
  Wu]{wang2021molecular}
Yingheng Wang, Yaosen Min, Erzhuo Shao, and Ji~Wu.
\newblock Molecular graph contrastive learning with parameterized explainable
  augmentations.
\newblock In \emph{2021 IEEE International Conference on Bioinformatics and
  Biomedicine (BIBM)}, pages 1558--1563. IEEE, 2021{\natexlab{a}}.

\bibitem[Wang et~al.(2021{\natexlab{b}})Wang, Chen, Min, and
  Wu]{wang2021molcloze}
Yingheng Wang, Xin Chen, Yaosen Min, and Ji~Wu.
\newblock Molcloze: a unified cloze-style self-supervised molecular structure
  learning model for chemical property prediction.
\newblock In \emph{2021 IEEE International Conference on Bioinformatics and
  Biomedicine (BIBM)}, pages 2896--2903. IEEE, 2021{\natexlab{b}}.

\bibitem[Kong et~al.(2022)Kong, Ricci, Guevarra, Neaton, Gomes, and
  Gregoire]{kong2022density}
Shufeng Kong, Francesco Ricci, Dan Guevarra, Jeffrey~B Neaton, Carla~P Gomes,
  and John~M Gregoire.
\newblock Density of states prediction for materials discovery via contrastive
  learning from probabilistic embeddings.
\newblock \emph{Nature communications}, 13\penalty0 (1):\penalty0 949, 2022.

\bibitem[Bai et~al.(2023)Bai, Du, Wang, Kong, Gregoire, and
  Gomes]{bai2023xtal2dos}
Junwen Bai, Yuanqi Du, Yingheng Wang, Shufeng Kong, John Gregoire, and Carla
  Gomes.
\newblock Xtal2dos: Attention-based crystal to sequence learning for density of
  states prediction.
\newblock \emph{arXiv preprint arXiv:2302.01486}, 2023.

\bibitem[Chen et~al.(2021)Chen, Bai, Ament, Zhao, Guevarra, Zhou, Selman, van
  Dover, Gregoire, and Gomes]{chen2021automating}
Di~Chen, Yiwei Bai, Sebastian Ament, Wenting Zhao, Dan Guevarra, Lan Zhou, Bart
  Selman, R~Bruce van Dover, John~M Gregoire, and Carla~P Gomes.
\newblock Automating crystal-structure phase mapping by combining deep learning
  with constraint reasoning.
\newblock \emph{Nature Machine Intelligence}, 3\penalty0 (9):\penalty0
  812--822, 2021.

\bibitem[Ament et~al.(2021)Ament, Amsler, Sutherland, Chang, Guevarra,
  Connolly, Gregoire, Thompson, Gomes, and van Dover]{ament2021autonomous}
Sebastian Ament, Maximilian Amsler, Duncan~R Sutherland, Ming-Chiang Chang, Dan
  Guevarra, Aine~B Connolly, John~M Gregoire, Michael~O Thompson, Carla~P
  Gomes, and R~Bruce van Dover.
\newblock Autonomous materials synthesis via hierarchical active learning of
  nonequilibrium phase diagrams.
\newblock \emph{Science Advances}, 7\penalty0 (51):\penalty0 eabg4930, 2021.

\bibitem[Degrave et~al.(2022)Degrave, Felici, Buchli, Neunert, Tracey,
  Carpanese, Ewalds, Hafner, Abdolmaleki, de~Las~Casas,
  et~al.]{degrave2022magnetic}
Jonas Degrave, Federico Felici, Jonas Buchli, Michael Neunert, Brendan Tracey,
  Francesco Carpanese, Timo Ewalds, Roland Hafner, Abbas Abdolmaleki, Diego
  de~Las~Casas, et~al.
\newblock Magnetic control of tokamak plasmas through deep reinforcement
  learning.
\newblock \emph{Nature}, 602\penalty0 (7897):\penalty0 414--419, 2022.

\bibitem[Blokhin and Villars(2019)]{blokhin2019materials}
Evgeny Blokhin and P~Villars.
\newblock Materials platform for data science: from big data towards materials
  genome.
\newblock 2019.

\bibitem[Gra{\v{z}}ulis et~al.(2012)Gra{\v{z}}ulis, Da{\v{s}}kevi{\v{c}},
  Merkys, Chateigner, Lutterotti, Quiros, Serebryanaya, Moeck, Downs, and
  Le~Bail]{gravzulis2012crystallography}
Saulius Gra{\v{z}}ulis, Adriana Da{\v{s}}kevi{\v{c}}, Andrius Merkys, Daniel
  Chateigner, Luca Lutterotti, Miguel Quiros, Nadezhda~R Serebryanaya, Peter
  Moeck, Robert~T Downs, and Armel Le~Bail.
\newblock Crystallography open database (cod): an open-access collection of
  crystal structures and platform for world-wide collaboration.
\newblock \emph{Nucleic acids research}, 40\penalty0 (D1):\penalty0 D420--D427,
  2012.

\bibitem[Saal et~al.(2013)Saal, Kirklin, Aykol, Meredig, and
  Wolverton]{saal2013materials}
James~E Saal, Scott Kirklin, Muratahan Aykol, Bryce Meredig, and Christopher
  Wolverton.
\newblock Materials design and discovery with high-throughput density
  functional theory: the open quantum materials database (oqmd).
\newblock \emph{Jom}, 65:\penalty0 1501--1509, 2013.

\bibitem[Hellenbrandt(2004)]{hellenbrandt2004inorganic}
Mariette Hellenbrandt.
\newblock The inorganic crystal structure database (icsd)—present and future.
\newblock \emph{Crystallography Reviews}, 10\penalty0 (1):\penalty0 17--22,
  2004.

\bibitem[Groom et~al.(2016)Groom, Bruno, Lightfoot, and Ward]{CSD}
Colin~R. Groom, Ian~J. Bruno, Matthew~P. Lightfoot, and Suzanna~C. Ward.
\newblock {The Cambridge Structural Database}.
\newblock \emph{Acta Crystallographica Section B}, 72\penalty0 (2):\penalty0
  171--179, Apr 2016.
\newblock \doi{10.1107/S2052520616003954}.
\newblock URL \url{https://doi.org/10.1107/S2052520616003954}.

\bibitem[Draxl and Scheffler(2019)]{NOMAD}
Claudia Draxl and Matthias Scheffler.
\newblock The nomad laboratory: from data sharing to artificial intelligence.
\newblock \emph{Journal of Physics: Materials}, 2\penalty0 (3):\penalty0
  036001, may 2019.
\newblock \doi{10.1088/2515-7639/ab13bb}.
\newblock URL \url{https://dx.doi.org/10.1088/2515-7639/ab13bb}.

\bibitem[Talirz et~al.(2020)Talirz, Kumbhar, Passaro, Yakutovich, Granata,
  Gargiulo, Borelli, Uhrin, Huber, Zoupanos, Adorf, Andersen, Sch{\"u}tt,
  Pignedoli, Passerone, VandeVondele, Schulthess, Smit, Pizzi, and Marzari]{MC}
Leopold Talirz, Snehal Kumbhar, Elsa Passaro, Aliaksandr~V. Yakutovich, Valeria
  Granata, Fernando Gargiulo, Marco Borelli, Martin Uhrin, Sebastiaan~P. Huber,
  Spyros Zoupanos, Carl~S. Adorf, Casper~Welzel Andersen, Ole Sch{\"u}tt,
  Carlo~A. Pignedoli, Daniele Passerone, Joost VandeVondele, Thomas~C.
  Schulthess, Berend Smit, Giovanni Pizzi, and Nicola Marzari.
\newblock Materials cloud, a platform for open computational science.
\newblock \emph{Scientific Data}, 7\penalty0 (1):\penalty0 299, Sep 2020.
\newblock ISSN 2052-4463.
\newblock \doi{10.1038/s41597-020-00637-5}.
\newblock URL \url{https://doi.org/10.1038/s41597-020-00637-5}.

\bibitem[Scheffler et~al.(2022)Scheffler, Aeschlimann, Albrecht, Bereau,
  Bungartz, Felser, Greiner, Gro{\ss}, Koch, Kremer, Nagel, Scheidgen,
  W{\"o}ll, and Draxl]{FAIRDataShare}
Matthias Scheffler, Martin Aeschlimann, Martin Albrecht, Tristan Bereau,
  Hans-Joachim Bungartz, Claudia Felser, Mark Greiner, Axel Gro{\ss},
  Christoph~T. Koch, Kurt Kremer, Wolfgang~E. Nagel, Markus Scheidgen, Christof
  W{\"o}ll, and Claudia Draxl.
\newblock Fair data enabling new horizons for materials research.
\newblock \emph{Nature}, 604\penalty0 (7907):\penalty0 635--642, Apr 2022.
\newblock ISSN 1476-4687.
\newblock \doi{10.1038/s41586-022-04501-x}.
\newblock URL \url{https://doi.org/10.1038/s41586-022-04501-x}.

\bibitem[Curtarolo et~al.(2012)Curtarolo, Setyawan, Hart, Jahnatek, Chepulskii,
  Taylor, Wang, Xue, Yang, Levy, Mehl, Stokes, Demchenko, and Morgan]{AFLOW}
Stefano Curtarolo, Wahyu Setyawan, Gus~L.W. Hart, Michal Jahnatek, Roman~V.
  Chepulskii, Richard~H. Taylor, Shidong Wang, Junkai Xue, Kesong Yang, Ohad
  Levy, Michael~J. Mehl, Harold~T. Stokes, Denis~O. Demchenko, and Dane Morgan.
\newblock Aflow: An automatic framework for high-throughput materials
  discovery.
\newblock \emph{Computational Materials Science}, 58:\penalty0 218--226, 2012.
\newblock ISSN 0927-0256.
\newblock \doi{https://doi.org/10.1016/j.commatsci.2012.02.005}.
\newblock URL
  \url{https://www.sciencedirect.com/science/article/pii/S0927025612000717}.

\end{thebibliography}

% \newpage
% \input{checklist}

\newpage
\appendix
\begin{center}
    {\Large \textbf{Appendix for M$^2$Hub}}
\end{center}

\tableofcontents
\renewcommand{\addcontentsline}[3]{\origaddcontentsline{#1}{#2}{#3}}

\section{Additional Related Work}
\label{sec:add_related}
As mentioned in Sec.~\ref{sec:conclusion}, our main focus in the current version is virtual screening, inverse design and molecular simulation while ignoring tasks related to experiments where problem formulation, dataset curation, evaluation and benchmarking are much more challenging. In this section, we introduce additional related work that leverage machine learning to assist in the experiment phase of materials discovery.

\subsection{Materials Synthesis}
In addition to designing materials computationally, new materials have to be synthesized with experiments. Even though computational methods have been deployed to simulate or predict material properties, it is much harder to predict how materials can be synthesized. Thus, it is a nontrivial problem to study material synthesis. Moreover, the synthesis process is challenging to predict as well. Material scientists have to post-process the synthesized material to identify the synthesized structures. Specifically, X-ray Diffraction (XRD) is a commonly used technology to detect crystal structures. ~\cite{chen2021automating} combines deep learning with reasoning module to incorporate physical constraints and identify crystal structure phase compositions from the experimental results with XRD patterns.

\subsection{Experiment Control with Machine Learning}
Experiment is an indispensable step to validate the properties of the materials. However, experiments are usually very expensive and even inaccessible in real scenarios. In addition, current experiment highly relies on human expert design which may be suboptimal. Therefore, automated experiment design becomes an urgent yet challenging problem. One way to improve the efficiency of experiment design is to leverage machine learning models for uncertainty estimation. Specifically, active learning can be utilized to construct a loop of decision and feedback. Machine learning models suggest the next experiment and the experiment provides feedback to improve machine learning models~\cite{ament2021autonomous}. Another promising direction is to leverage reinforcement learning methods which have an agent attempting to achieve some goal by the feedback provided by the environment~\cite{degrave2022magnetic}. 

\section{Additional Data Sources}
\label{sec:add_data}

In addition to the datasets included in our current version (Sec.~\ref{sec:overview}), there are other available data sources which may be used in different purposes and we will consider to add in our future version. 

\begin{itemize}
\item Materials Platform for Data Science (MPDS)~\cite{blokhin2019materials} (no license) is a data repository that collects experimental and computational materials data through data mining from the scientific publications. There, around half a million articles were manually processed and systematized, covering a broad spectrum of physical sciences, such as physics, chemistry, materials science, environmental science, engineering, and geology. 
\item Crystallography Open Database (COD)~\cite{gravzulis2012crystallography} (license: CC0 1.0): COD is a open-access database containing crystallographic data on inorganic and organic compounds. It includes experimentally determined crystal structures along with associated metadata for organic, inorganic, metal-organic compounds and minerals.
\item Open Quantum Materials Database (OQMD)~\cite{saal2013materials} (license: CC-BY 4.0): OQMD is a database that focuses on quantum-mechanical calculations of materials properties. It contains calculated data on crystal structures, electronic structures, formation energies, and other material properties for a wide range of inorganic compounds.
\item Inorganic Crystal Structure Database (ICSD)~\cite{hellenbrandt2004inorganic} (commercial): ICSD is a comprehensive database that compiles > 281,000 experimentally determined crystal structures of inorganic compounds. To ensure the high quality of structures in ICSD, a structure has to be fully characterized and passed thorough quality checks by its expert editorial team before inclusion. The information in ICSD is updated biannually.
\item Cambridge Structural Database (CSD)~\cite{CSD} (commercial): CSD contains over 1.1M accurate 3D experimentally crystalized structures with data from X-ray and neutron diffraction analyses. It contains diverse types of organic crystal structure (drug, pigment, etc.) and metal-organic crystals (transition metal complex, metal-organic framework, etc.).
\item Novel Materials Discovery (NOMAD)~\cite{NOMAD} (license: CC-BY 4.0): NOMAD is a data management platform for materials science data where users can share data freely. Here, NOMAD is a web-application and database that allows to centrally publish data. But you can also use the its utilities to build your own local database.
\item Materials Cloud (MC)~\cite{MC} (license: CC0 1.0): MC is built to enable the seamless sharing and dissemination of resources in computational materials science, offering educational, research, and archiving tools; simulation software and services; and curated and raw data. These underpin published results and empower data-based discovery, compliant with data management plans and the FAIR principles~\cite{FAIRDataShare}. In addition to database, MC also provides lectures for computational materials science, various visualization and simulation tools.
\item AFLOW~\cite{AFLOW} (MIT license) Similar to MC, AFLOW is a composite platform includes materials database, search and visualization, simulation, and machine learning models.
\end{itemize}

\section{Additional Experimental Results}
\label{sec:add_results}

In addition to our benchmark results, two popular benchmarking datasets, OC20 and QM9 have been extensively tested in previous work. We directly take the experimental results from~\cite{liaoequiformer} for OC20 (Table.\ref{tab:oc20}) and~\cite{LEFTNet} for QM9 (Table.\ref{tab:qm9}) as a reference on performance of existing approaches. Note that there are two commonly used data splits for QM9 in previous literature and they are both reported.

\begin{table}
\begin{center}
\caption{Benchmark on machine learning forcefield (OC20 IS2RE test set) (results taken from~\cite{liaoequiformer}). (The best results are \textbf{bolded}.)}
\begin{adjustbox}{max width=\textwidth}
\begin{tabular}{cccccc|ccccccc}
\toprule
 & \multicolumn{5}{c|}{Energy MAE} & \multicolumn{5}{c}{EwT} \\\hline
Methods & ID & OOD Ads & OOD Cat & OOD Both & Average & ID & OOD Ads & OOD Cat & OOD Both & Average \\\hline 
CGCNN & 0.6149 & 0.9155 & 0.6219 &0.8511& 0.7509 & 3.40 &1.93 &3.10 &2.00 &2.61  \\
SchNet & 0.6387 &0.7342& 0.6616& 0.7037& 0.6846& 2.96& 2.33& 2.94& 2.21& 2.61\\
% PaiNN~\cite{schutt2021equivariant} &  &  &  & \\
% SphereNet~\cite{liu2022spherical} &  &  &  & \\
DimeNet++ & 0.5621 & 0.7252 & 0.5756 & 0.6613 & 0.6311& 4.25& 2.07& 4.10& 2.41& 3.21\\
% GemNet &  &  &  &  \\
Equiformer & \textbf{0.5037} & \textbf{0.6881} & \textbf{0.5213} &\textbf{0.6301}& \textbf{0.5858}& \textbf{5.14}& \textbf{2.41}& \textbf{4.67}& \textbf{2.69}& \textbf{3.73}  \\
% LEFTNet~\cite{LEFTNet} & &  &  &  \\
\bottomrule
\end{tabular}
\label{tab:oc20}
\end{adjustbox}
\end{center}
\end{table}

\begin{table*}[t]
  \centering
  \caption{Benchmark on molecular property prediction (QM9) (results taken from~\cite{LEFTNet}). (The best results are \textbf{bolded}.)}
  \resizebox{0.95\textwidth}{!}{
  \begin{tabular}{l c c c c c c c c c c c c}
  \toprule
    Task & $\alpha$ & $\Delta \varepsilon$ & $\varepsilon_{\mathrm{HOMO}}$ & $\varepsilon_{\mathrm{LUMO}}$ & $\mu$ & $C_{\nu}$ & $G$ & $H$ & $R^2$ & $U$ & $U_0$ & ZPVE \\
    Units & bohr$^3$ & meV & meV & meV & D & cal/mol K & meV & meV & bohr$^3$ & meV & meV & meV \\
    \midrule
    EGNN & .071  & 48 & 29 & 25 & .029 & .031 & 12 & 12 & \textbf{.106} & 12 & 11 & 1.55  \\
    Equiformer & .056 & \textbf{33} & \textbf{17} & \textbf{16} & .014 & .025 & 10 &10 &.227 &11 &10 &1.32\\
    LEFTNet & \textbf{.048} & 40 & 24 & 18 & \textbf{.012} & \textbf{.023}
 & \textbf{7} & \textbf{6} & .109 & \textbf{7} & \textbf{6} & 1.33\\
    \midrule
    SchNet & .235 & 63 & 41 & 34 & .033 & .033 & 14 & 14 & .073 & 19 & 14 & 1.70 \\
    DimeNet++ & .044 & 33 & 25 & 20 & .030 & .023 & 8 & 7 & .331 & 6 & 6 & 1.21  \\
    LEFTNet & .039 & 39 & \textbf{23} & \textbf{18} & \textbf{.011} & .022 & \textbf{6} & \textbf{5} & .094 & \textbf{5} & \textbf{5} & 1.19 \\
    \bottomrule
  \end{tabular}}
  \label{tab:qm9}
\end{table*}

\section{Dataset Statistics}

In Table~\ref{tab:stats}, we report the statistics of each dataset with number of samples and number of atoms in each material.

\begin{table}[ht!]
\begin{center}
\caption{Dataset statistics (number of samples 
in each dataset and size of systems in each dataset.}
\label{tab:stats}
\begin{adjustbox}{max width=\textwidth}
\begin{tabular}{cccccccccccccc}
\toprule
Datasets & dft2d & edos & pdos & qmof bandgap & omdb bandgap & dielectric & log\_gvrh & log\_kvrh & e\_form & mp bandgap & is\_metal & perovskites & phonons \\\hline 
number of samples & 636 & 55,659 & 14,244 & 20,425 & 12,500 & 4,764 & 10,987 & 10,987 & 132,752 & 106,113 & 106,113 & 18,928 & 1,265 \\
number of atoms & 7.19$\pm$ 4.35 & 10.08$\pm$9.06 & 7.23$\pm$ 5.46 & 113.67$\pm$68.86 & 82.29$\pm$ 26.55 & 16.89$\pm$14.67 & 8.63$\pm$8.66 & 8.63$\pm$8.66 & 29.15$\pm$30.1 & 30.02$\pm$29.94 & 30.02$\pm$29.94 & 5.00$\pm$0.00 & $7.63\pm 3.74$ \\
\bottomrule
\end{tabular}
\end{adjustbox}
\end{center}
\end{table}

\section{Oracle Function Details}
\label{sec:oracle}

\paragraph{FP-based oracle function} This method generates desired properties for any specific input materials. The core concept is SCM/MagPie featurization and machine learning prediction. The oracle function first reads CIF files to extract the structures or compositions of the materials. Then, the function preprocess and prepares the appropriate data format based on the task at hand. Before training, the materials data are transformed by the Sine Coulomb Matrix and MagPie featurization algorithms to convert the raw data into a form that can be used by the machine learning model. After that, a standard machine learning pipeline is built to predict the target property for these materials. The pipeline uses random forest as the machine learning model. Finally, the method saves the predictions and returns them for user convenience.
\paragraph{Structure-based oracle function} This method is used to match a given film with a list of substrates. The function first reads the input film and substrates to get their structures. Then, the \emph{SubstrateAnalyzer} from \emph{pymatgen} \cite{ong2013python} is called to calculate possible matches between the film and each substrate. It finds the best matches (with the smallest matching area) for each orientation of the substrate. Finally, all the match information is stored and returned for users. The match information includes the substrate's formula, orientations of the substrate and film, the matching area, and optionally, the elastic energy and strain. This substrate matching process can be useful in thin film deposition processes, where you want to match the crystal structure of a thin film material to a substrate material to ensure good adhesion and minimize defects.

\begin{table}[ht!]
\begin{center}
\caption{Hyparameters for benchmarked machine learning models.}
\label{tab:hyperparam}
\begin{adjustbox}{max width=\textwidth}
\begin{tabular}{ccccccccc}
\toprule
& CGCNN & ALIGNN & SchNet & EGNN & DimeNet++ & GemNet & Equiformer & LEFTNet\\\midrule
cutoff & 6.0 & 8.0 & 6.0 & 6.0 & 6.0 & 6.0 & 6.0 & 6.0\\
max\_neighbors & N/A & 12 & N/A & N/A & N/A & 50 & 500 & N/A\\
num\_layers & 5 & 3 & 6 & 4 & 3 & 3 & 6 & 4\\
hidden\_dimension & 256 & 128 & 128 & 128 & 192 & 128 & 512 & 128 \\
learning rate & 1e-4 & 1e-3 & 1e-4 & 1e-4 & 1e-4 & 5e-4 & 1e-3 & 5e-4 \\
optimizer & AdamW & AdamW & AdamW & Adam & AdamW & AdamW & AdamW & AdamW \\
scheduler & N/A & N/A & N/A & N/A & N/A & ReduceLROnPlateau & Cosine & N/A\\
training epochs & 500 & 500 & 500 & 500 & 500 & 500 & 500 & 500 \\
\bottomrule
\end{tabular}
\end{adjustbox}
\end{center}
\end{table}

\section{Experimental Details}
\label{sec:exp_detail}

\subsection{Hyperparameters}

We report the general hyperparamters shared across models in Table~\ref{tab:hyperparam}. For model-specific parameters, we report in \url{https://github.com/yuanqidu/M2Hub/config}.

\end{document}